\documentstyle[twocolumn,seceq,epsfig]{jpsj}

\def\runtitle{
Nonvanishing Local Moment in Triplet Superconductors
}

\def\runauthor{Mikito {\sc Koga} and Masashige {\sc Matsumoto}}

\title{
Nonvanishing Local Moment in Triplet Superconductors
}

\author{Mikito {\sc Koga} and Masashige {\sc Matsumoto}$^1$}

\inst{
Department of Physics, Faculty of Education, Shizuoka University, 836 Oya,
Shizuoka 422-8529, Japan
$^1$Department of Physics, Faculty of Science, Shizuoka University, 836 Oya,
Shizuoka 422-8529, Japan \\
}

\recdate{\today}

\abst{
The Kondo effect in a $p_x + {\rm i} p_y$-wave superconductor is studied by
applying the Wilson's numerical renormalization group method.
In this type of superconductor with a full energy gap like a $s$-wave one,
the ground state is always a spin doublet,
while a local spin is shrunk by the Kondo effect.
The calculated magnetic susceptibility indicates that
the spin of the ground state is generated by the orbital effect of
the $p_x + {\rm i} p_y$-wave Cooper pairs.
The effect of spin polarization of the triplet superconductor is also discussed.
}

\kword{
Kondo effect, unconventional superconductivity, triplet Cooper pair,
numerical renormalization group
}

\begin{document}
\sloppy
\maketitle
\newcommand{\br}{{\mbox{\boldmath$r$}}}
\newcommand{\bk}{{\mbox{\boldmath$k$}}}
\newcommand{\sk}{{\mbox{\footnotesize $k$}}}
\newcommand{\bsk}{{\mbox{\footnotesize \boldmath$k$}}}
\newcommand{\bS}{{\mbox{\boldmath$S$}}}
\newcommand{\bd}{{\mbox{\boldmath$d$}}}
\newcommand{\bsigma}{{\mbox{\boldmath$\sigma$}}}
\newcommand{\fig}[1]
{\vspace{24pt}
\begin{center}
\fbox{\rule{0cm}{#1}\hspace{7cm}}
\end{center}}

\section{Introduction}
The Kondo effect is strongly suppressed by existence of an energy gap at the Fermi
energy, since infinitesimal excitations are important.
In fact, two energy scales, the Kondo temperature $T_{\rm K}$ and the energy gap
$\Delta$, compete with each other.
In a standard BCS ($s$-wave) superconductor, the competition
is clearly seen in $T_{\rm K} / \Delta$ dependence of a bound state energy
below the energy gap.
\cite{Satori,Sakai93,Yoshioka}
The ground state is a spin doublet for $T_{\rm K} / \Delta \ll 1$ and it is
a spin singlet for $T_{\rm K} / \Delta \gg 1$.
The interchange of ground states occurs at around $T_{\rm K} / \Delta = 0.3$.
\par

According to the recent nuclear magnetic resonance measurement performed
in Li-substituted YBa$_2$Cu$_3$O$_{7-\delta}$,
\cite{Bobroff}
the local spin susceptibility indicates the reduction of the Kondo effect due to
opening of the superconducting energy gap below the superconducting transition temperature.
This is supported by the recent theoretical study on the Kondo effect in a
$d_{x^2-y^2}$-wave superconductor.
\cite{Simon}
It seems that the interchange of spin-doublet and spin-singlet ground states also
occurs in such singlet superconductors as the $s$-wave and $d$-wave.
\par

In this paper, we study the Kondo effect in triplet superconductors, using the numerical
renormalization group (NRG) method.
\cite{Wilson,Sakai92}
The simplest form for the NRG calculation is a $p_x + {\rm i} p_y$-wave pairing type
with a full energy gap like the $s$-wave.
One of the $p_x + {\rm i} p_y$-wave pairing states is given by an odd vectorial function
$\bd = \hat{z} (k_x + {\rm i} k_y)$ or is expressed more explicitly with a gap function
$\Delta_{\sigma \sigma'}(\bk) = \Delta_{\uparrow \downarrow} (k_x + {\rm i} k_y)$.
\cite{Sigrist2}
A possibility of this state has been discussed for a triplet superconductor
Sr$_2$RuO$_4$ in which the time-reversal symmetry is broken.
\cite{Maeno,Sigrist}
We also study the case of a spin-polarized superconducting state represented by both
$\Delta_{\uparrow \uparrow} (k_x + {\rm i} k_y)$ and
$\Delta_{\downarrow \downarrow} (k_x + {\rm i} k_y)$.
This state is a possible candidate proposed for a ferromagnetic superconductor recently
discovered in UGe$_2$,
\cite{Saxena,Machida}
where such a triplet superconductivity is expected to be realized under high
pressure environment.
This is also expected in other ferromagnetic metals such as ZrZn$_2$
\cite{Peiderer}
and URhGe.
\cite{Aoki}
\par

Since the Cooper pair of the $p_x + {\rm i} p_y$-wave state has a net angular momentum $l=1$,
it is formed by conduction electrons having $l=0$ and $l=1$ angular momenta.
A local spin is coupled directly to the $l = 0$ conduction electrons.
When the superconducting state is set on, the $l = 1$ conduction electrons also
participate in the Kondo effect.
In this case, the ground state is a spin doublet for all values of $T_{\rm K} / \Delta$,
\cite{Matsumoto}
which is different from the $s$-wave case.
In order to elucidate magnetic properties of this new type of spin-doublet ground state,
we calculate magnetic susceptibilities of the impurity.
In the Kondo problems, two kinds of magnetic susceptibilities are often discussed:
one is a local spin susceptibility $\chi_{\rm local}$ and the other is called an impurity
susceptibility $\chi_{\rm imp}$.
The size of the local spin is reduced by the Kondo effect as $T_{\rm K} / \Delta$ increases,
which is reflected in the zero temperature limit of $T\chi_{\rm local}$.
In such gapped systems as superconductors, it is qualitatively different from $T\chi_{\rm imp}$.
The difference is also found in semi-conducting systems where a pseudo-gap exists at the
Fermi energy.
\cite{Withoff,Chen,Gonzalez}
It is more evident in the fully gapped system we discuss here.
\par

We also discuss the spin effect of Cooper pairs in the triplet superconductors
by introducing spin polarization
($|\Delta_{\uparrow \uparrow}| - |\Delta_{\downarrow \downarrow}|$).
When either $\Delta_{\uparrow \uparrow}$ or $\Delta_{\downarrow \downarrow}$ vanishes,
the infinitesimal excitations are the most relevant and the Kondo effect becomes inactive.
In this case, the spin direction of low-energy excitations is locked by the zero order parameter,
and the local spin behaves as a classical spin.
\par

This paper is organized as follows.
In \S2, we present the Kondo model for the $p_x + i p_y$ superconductivity and the
corresponding NRG Hamiltonian.
In \S3, we show the NRG results for spin-unpolarized and spin-polarized superconducting
states separately.
Concluding remarks are given in \S4.
\par

\section{Model Hamiltonian}
Let us begin with the following Hamiltonian for the spin-unpolarized triplet Cooper pair
represented by $\Delta_{\uparrow \downarrow} = \Delta_{\downarrow \uparrow} = \Delta$:
\cite{Matsumoto}
\begin{eqnarray}
&& H = H_{\rm kin} + H_\Delta + H_{\rm imp}, \\
&& H_{\rm kin}
   = \int_{-1}^1 {\rm d}k k \sum_{l\sigma} a_{\sk l \sigma}^\dagger a_{\sk l \sigma},
\label{eqn:2} \\
&& H_\Delta = \int_{-1}^1 {\rm d}k \sum_l (-1)^{1-l}
   \left( {\rm i} \Delta a_{\sk l\uparrow}^\dagger a_{\sk,-l+1,\downarrow}^\dagger
   + {\rm H. c.} \right),
\nonumber \\
&&
\label{eqn:3} \\
&& H_{\rm imp} = - \sum_{\sigma\sigma'}
   J \bS \cdot \bsigma_{\sigma\sigma'} f_{00\sigma}^\dagger f_{00\sigma'}, \\
&& f_{00\sigma} = \frac{1}{\sqrt{2}} \int_{-1}^1 {\rm d}k a_{\sk,l=0,\sigma}.
\end{eqnarray}
\noindent
Here $a_{kl\sigma}$ is an annihilation operator of the conduction electrons,
where the subscripts $k$, $l$, and $\sigma$
represent wave number renormalized by $k_{\rm F}$, angular momentum and spin, respectively.
$J$ ($<0$) is an antiferromagnetic exchange coupling constant.
The dispersion relation of the conduction electrons
has been linearized for each angular momentum $l$ in eq.~(\ref{eqn:2}).
The total angular momentum of the $p_x + {\rm i} p_y$-wave Cooper pair is equal to one,
which is given by the coupling of two angular momenta, $l$ and $-l + 1$
($l = 0, \pm 1, \pm 2, \cdots$), as shown in eq.~(\ref{eqn:3}).
The local spin is coupled directly to the $l = 0$ electrons.
Therefore the relevant $p_x + {\rm i} p_y$-wave Cooper pair in the Kondo effect consists of
the $l=0$ and $l=1$ conduction electrons.
In the normal metallic state ($\Delta=0$),
the $l=0$ conduction electrons screen the local spin,
while the $l=1$ conduction electrons are completely decoupled from the local spin.
In the superconducting state ($\Delta \neq 0$),
the $l=1$ conduction electrons participate in the Kondo effect.
\par

As we have derived in our previous paper,
\cite{Matsumoto}
the NRG Hamiltonian for the $p_x + {\rm i} p_y$-wave is given in the following recursion relation:
\begin{eqnarray}
&&H_{N+1} = \Lambda^{1/2}H_N + \sum_{\tau\sigma}
  \Bigl[
    \varepsilon_N ( c_{N+1,\tau\sigma}^\dagger c_{N\tau\sigma} + {\rm H.c.} ) \cr
&&~~~~~~~~~~~~~~~
     +(-1)^N\Lambda^{N/2}\tau\tilde{\Delta}
        c_{N+1,\tau\sigma}^\dagger c_{N+1,\tau\sigma}
  \Bigr],
\label{eqn:6}
\\
&&H_0 =
  \Bigl[ \frac{1}{2}
    \sum_{\tau\tau'\sigma\sigma'}
      ( - \tilde{J}) \mbox{\boldmath$S$} \cdot \mbox{\boldmath$\sigma$}_{\sigma\sigma'}
        c_{0\tau\sigma}^\dagger c_{0\tau'\sigma'} \cr
&&~~~~~~~~~~~~~~~ -\sum_{\tau\sigma} \tau\tilde{\Delta} c_{0\tau\sigma}^\dagger c_{0\tau\sigma}
  \Bigr] \Lambda^{-1/2}.
\label{eqn:7}
\end{eqnarray}
$\tilde{J}$ and $\tilde{\Delta}$ are the effective exchange coupling $J$
and the superconducting energy gap $\Delta$ normalized by $(1 + \Lambda^{-1})/2$.
$\Lambda$ is a logarithmic discretization parameter and it is taken to be $\Lambda = 3$
throughout this paper.
$c_{N\tau\sigma}$ is an operator of the NRG fermion quasiparticle in the $N$-th shell.
The subscript $\tau=\pm$ represents the two channels
which are constructed by the $l=0$ and $l=1$ orbitals.
The Kondo temperature is defined by $T_{\rm K} = |J|^{1/2} \exp (-1/|J|)$.
\par

In the same manner, we can derive the NRG Hamiltonian for describing the Kondo effect in the
spin-polarized $p_x + {\rm i} p_y$-wave state represented by
$\Delta_{\uparrow \uparrow}$ and $\Delta_{\downarrow \downarrow}$.
Since the BCS interaction corresponding to eq.~(\ref{eqn:3}) is given by
\begin{equation}
   H_\Delta = \int_{-1}^1 {\rm d}k \sum_{l \sigma} (-1)^{1-l}
   \left( {\rm i} \Delta_{\sigma \sigma} a_{\sk l\sigma}^\dagger a_{\sk,-l+1,\sigma}^\dagger
   + {\rm H. c.} \right),
\end{equation}
we only put spin dependence into the energy gap as
$\tilde{\Delta} \rightarrow \tilde{\Delta}_{\sigma \sigma}$
in eqs.~(\ref{eqn:6}) and (\ref{eqn:7}).
\par

Next we describe how to calculate magnetic susceptibilities.
There are two kinds of them.
One is a local impurity susceptibility defined by
\begin{eqnarray}
\chi_{\rm local}(T) &=& g\mu_{\rm B} \frac{\langle S_z \rangle}{h}|_{h\rightarrow 0}, \\
\langle S_z \rangle &=& \frac{ {\rm Tr} \left[S_z {\rm exp}(-\bar{\beta} H_N)\right]}
                  { {\rm Tr} \left[{\rm exp}(-\bar{\beta} H_N)\right]}.
\end{eqnarray}
Here $S_z$ is the local spin operator, and $\bar{\beta}\sim 1$ is taken
to reduce cutoff dependence in the NRG calculation.
The temperature $T$ is defined by
$\bar{\beta} T = [(1+\Lambda^{-1})/2] \Lambda^{-(N-1)/2}$.
\cite{Wilson,Sakai92}
A small magnetic field $h$ is applied only at the local spin site.
In practice, the impurity part of the Hamiltonian $H_0$ in eq. (\ref{eqn:7}) is replaced
by $H_0 - g \mu_{\rm B} S_z h \Lambda^{-1/2}$ in the NRG calculation.
The other susceptibility is defined by
\begin{equation}
\chi_{\rm imp}(T) = \chi(T) - \chi_0(T),
\end{equation}
where $\chi$ ($\chi_0$) is a magnetic susceptibility of the conduction electrons
including (in the absence of) the local spin.
\par

\section{Results}
In the remaining parts of this paper, we show the NRG results for the spin-unpolarized
($\Delta_{\uparrow \downarrow} = \Delta_{\downarrow \uparrow}$,
$\Delta_{\uparrow \uparrow} = \Delta_{\downarrow \downarrow} = 0$)
and spin-polarized  ($\Delta_{\uparrow \uparrow} \ne \Delta_{\downarrow \downarrow}$,
$\Delta_{\uparrow \downarrow} = \Delta_{\downarrow \uparrow} = 0$)
superconducting states.
\par

\subsection{Spin-unpolarized superconductivity}

\begin{figure}[t]
\begin{center}
\psfig{figure=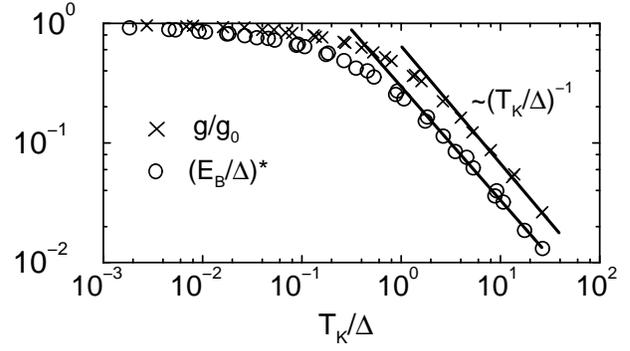,width=8.0cm}
\end{center}
\caption{
$T_{\rm K}/\Delta$ dependence of the bound state energy $(E_{\rm B}/\Delta)^*$
and the effective $g$-factor for the spin-unpolarized superconductivity.
$g_0$ is a bare $g$-factor of the local spin.
The lowest-lying $\sim 1200$ states are kept at each renormalization step in the
NRG calculation.
}
\label{fig:1}
\end{figure}
\begin{figure}[t]
\begin{center}
\psfig{figure=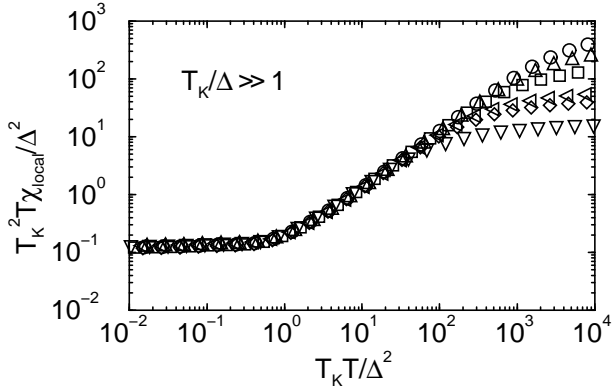,width=8.0cm}
\end{center}
\caption{
Temperature dependence of $T\chi_{\rm local}$ in the strong coupling region.
The unit is $(g_0 \mu_{\rm B})^2$, where $g_0$ is a bare $g$-factor of the local spin.
The data for $T_{\rm K} / \Delta > 8$ are plotted here:
the circles ($\tilde{\Delta} = 1.5 \Delta = 5 \times 10^{-5}$),
squares ($1 \times 10^{-4}$), and diamonds ($2 \times 10^{-4}$) are for the fixed
$T_{\rm K} = 1.76 \times 10^{-3}$;
the up triangle ($\tilde{\Delta} = 2 \times 10^{-5}$), left triangle
($5 \times 10^{-5}$), and down triangle ($1 \times 10^{-4}$) symbols are for
the fixed $T_{\rm K} = 5.34 \times 10^{-4}$.
In the middle region, $\chi_{\rm local}$ shows a constant value ($\propto T_{\rm K}^{-1}$).
The zero temperature limit gives the effective $g$-factor ($\propto \Delta / T_{\rm K}$)
in Fig.~\ref{fig:1}.
}
\label{fig:2}
\end{figure}

In our previous study,
\cite{Matsumoto}
we found that the ground state is always a spin doublet ($S = 1/2$)
in all the $T_{\rm K}/\Delta$ region.
The first excited state (bound state) is a particle-hole doublet with no spin ($S = 0$).
The ratio of the bound state energy and the superconducting energy gap converges
after the renormalization process.
We express the convergent value with $(E_{\rm B}/\Delta)^*$.
It is scaled by $T_{\rm K}/\Delta$, as shown in Fig~\ref{fig:1}.
As $T_{\rm K} / \Delta$ increases, the bound state ($S = 0$) energy approaches
that of the spin-doublet ground state ($S = 1/2$).
It implies that the local spin is quenched at $T_{\rm K} / \Delta \rightarrow \infty$.
The screening of the local spin is clearly found in the zero temperature limit of
$T\chi_{\rm local}$, exhibiting the Curie law.
Therefore this value corresponds to an effective $g$-factor of the local spin.
In Fig.~\ref{fig:1}, the bound state energy and the $g$-factor
exhibit similar $T_{\rm K}/\Delta$ dependence.
In the strong coupling region ($T_{\rm K}/\Delta \gg 1$),
both of them behave as $\sim \Delta / T_{\rm K}$.
In addition, the temperature dependence of $T \chi_{\rm local}$ for $T < T_{\rm K}$ is scaled
nicely as shown in Fig.~\ref{fig:2}.
This can be expressed by the following equation:
\begin{equation}
{T\chi_{\rm local} \over (g_0 \mu_{\rm B})^2} = \left({\Delta \over T_{\rm K}} \right)^2
f \left({T_{\rm K} T \over \Delta^2} \right),
\end{equation}
where $g_0$ is a bare $g$-factor of the local spin and the function $f(x)$ behaves as
\begin{eqnarray}
&& f(x) \propto x ~~ (1 < x < (T_{\rm K} / \Delta)^2), \\
&& f(x) \simeq 10^{-1} ~~ (0 < x < 1).
\end{eqnarray}
The constant $\chi_{\rm local} \sim T_{\rm K}^{-1}$ in the middle region corresponds
to the low temperature behavior in the metallic case ($\Delta = 0$).
On the other hand, at high temperatures ($T > T_{\rm K}$), $T\chi_{\rm local}$ is
scaled by the single energy $T_{\rm K}$ as expected by the usual Kondo effect,
and the energy gap $\Delta$ can be neglected.
As $T_{\rm K} / \Delta$ decreases, the middle region of $T\chi_{\rm local}$ is narrower.
For the weak coupling limit ($T_{\rm K} / \Delta \ll 1$), $T\chi_{\rm local}$ is not
reduced so much at high temperatures ($T > \Delta$), and it becomes constant near to $1/4$
for $T < \Delta$, corresponding to the value for a free $S = 1/2$ local spin.
Thus the competition between the Kondo effect and the energy gap is observed clearly by the
zero temperature limit of $T\chi_{\rm local}$ (the effective $g$-factor) and its temperature
dependence for $T_{\rm K} / \Delta \gg 1$.
\par

\begin{figure}[t]
\begin{center}
\psfig{figure=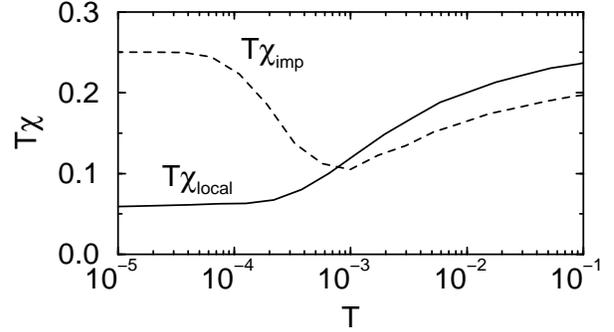,width=8.0cm}
\end{center}
\caption{
Temperature dependence of $T\chi_{\rm imp}$ and $T\chi_{\rm local}$
for the spin-unpolarized superconductivity.
$T_{\rm K}$ and $\tilde{\Delta}$($= 1.5 \Delta$) are chosen as
$T_{\rm K}=5.34\times 10^{-4}$ and $\tilde{\Delta}=10^{-3}$, respectively.
The lowest-lying $\sim 1200$ states are kept at each renormalization step in the
NRG calculation.
}
\label{fig:3}
\end{figure}

Next we discuss the two kinds of magnetic susceptibility $\chi_{\rm local}$ and
$\chi_{\rm imp}$.
They show the different temperature dependence as seen in Fig.~\ref{fig:3}.
At higher temperatures, we can see that both exhibit similar temperature dependence.
For $T\chi_{\rm imp}$, it upturns at sufficiently low temperatures
and deviates from $T\chi_{\rm local}$.
It saturates at low temperatures and takes 1/4 which is the same value with
$T\chi_{\rm local}$ for a free $S = 1/2$ local spin.
On the other hand, $T\chi_{\rm local}$ in Fig.~\ref{fig:3} is reduced by the Kondo effect.
What does this difference mean?
Let us discuss this point in the strong coupling limit ($T_{\rm K}/\Delta\rightarrow\infty$)
where $T \chi_{\rm local} \rightarrow 0$ at low temperatures.
Although the ground state keeps a spin doublet,
the effective $g$-factor of the local spin is strongly reduced
by the Kondo effect (see Fig. \ref{fig:1}).
This implies that the $l=0$ conduction electrons quench the local spin by forming a Kondo singlet.
In order to gain the superconducting condensation energy,
the $l=1$ conduction electrons form Cooper pairs with the $l=0$ electrons.
As a result,
the $l = 1$ electrons are connected with the Kondo singlet.
This concludes that the spin of the ground state is generated by the $l=1$ conduction electrons
in the strong coupling limit.
For $\chi_{\rm local}$,
the $l=1$ conduction electrons do not respond to the magnetic field directly,
since it is applied only at the impurity.
On the other hand, their magnetic response is reflected in $\chi_{\rm imp}$.
Thus $\chi_{\rm imp}$ represents the magnetic response
to an external field applied not only at the local impurity
but also to the conduction electrons.
The qualitative difference between $\chi_{\rm local}$ and $\chi_{\rm imp}$ is obvious in
the gapped systems.
\par

In Fig.~\ref{fig:3}, we note
that the upturn in $T\chi_{\rm imp}$ takes place at around $T=\Delta$.
For $T>\Delta$, the energy gap does not affect $\chi_{\rm imp}$,
so that $T\chi_{\rm imp}$ has almost the same temperature dependence with $T\chi_{\rm local}$.
Therefore we conclude that the different behavior between $\chi_{\rm local}$ and
$\chi_{\rm imp}$ for $T < \Delta$ is due to the finite energy gap in the density of
conduction electron states.
This should be compared with a pseudo-gap case where such a difference is also found.
\cite{Withoff,Chen,Gonzalez}
We note that infinitesimal excitations exist in this case, while we treat the full gap
systems.
If the density of states at the Fermi energy ($\varepsilon = 0$) is given by
$\rho \propto |\varepsilon|^r$, both $T \chi_{\rm local}$($T \rightarrow 0$) and
$T \chi_{\rm imp}$($T \rightarrow 0$) depend on $r$ strongly.
The former is extremely small, while the latter takes finite values between $0$ and $1/4$.
In general, the difference of the two susceptibilities is due to the existence of an energy gap.
Our study on the full-gap superconducting state shows this fact more evidently.
\par

\subsection{Spin-polarized superconductivity}

\begin{figure}[t]
\begin{center}
\psfig{figure=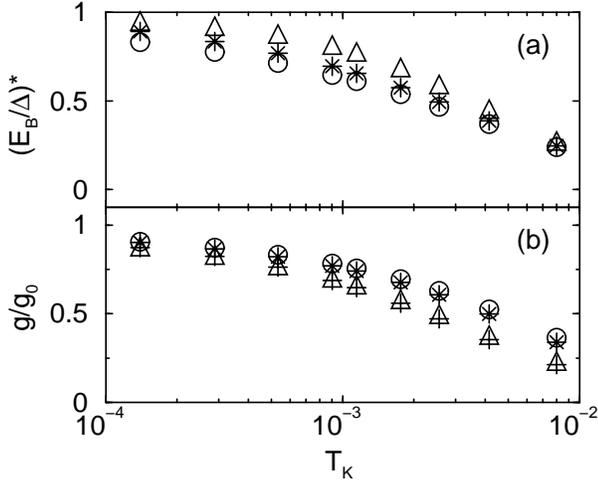,width=8.0cm}
\end{center}
\caption{
(a)~$T_{\rm K}$ dependence of the bound state energy
$(E_{\rm B}/\Delta)^* = (E_{\rm B} / \Delta_{\downarrow \downarrow})^*$
and (b)~$T_{\rm K}$ dependence of the effective $g$-factor.
$g_0$ is a bare value of the $g$-factor of the local spin.
In both figures, $\tilde{\Delta}_{\uparrow \uparrow} = 0.01$ is taken.
The circle, star, triangle and plus symbols represent the results
for fixed $\tilde{\Delta}$($= 1.5 \Delta$) $= 0.01$, $0.008$, $0.001$ and $0$, respectively.
The ground state is always a spin doublet
and the energies of the bound state are measured from those of the ground state.
The first excited state is a particle-hole doublet ($S = 0$).
The lowest-lying $\sim 500$ states are kept at each renormalization step in the
NRG calculation here.
}
\label{fig:4}
\end{figure}
The spin-polarized superconducting state is represented by the spin dependent
order parameters $\Delta_{\uparrow \uparrow}$ and $\Delta_{\downarrow \downarrow}$.
For $\Delta_{\uparrow \uparrow} = \Delta_{\downarrow \downarrow}$, we obtain the
same results as in the unpolarized spin case.
The polarization $(|\Delta_{\uparrow \uparrow}| - |\Delta_{\downarrow \downarrow}|)$
suppress the strength of the Kondo effect, so that it increases the bound state
($S = 0$) energy.
In Fig.~\ref{fig:4}(a), the effect of the difference between $\Delta_{\uparrow \uparrow}$
and $\Delta_{\downarrow \downarrow}$ is found as the increase of the bound state
energy for each values of $T_{\rm K}$ when $T_{\rm K}$ is smaller than
$\Delta_{\uparrow \uparrow}$.
As expected, the $T_{\rm K}$ dependence is close to that for the unpolarized case
($\Delta_{\uparrow \uparrow} = \Delta_{\downarrow \downarrow}$) if
$T_{\rm K} \gg \max (|\Delta_{\uparrow \uparrow}|, |\Delta_{\downarrow \downarrow}|)$.
We also calculate $\chi_{\rm local}$ for each value of
$\Delta_{\downarrow \downarrow}$, keeping $\Delta_{\uparrow \uparrow}$ fixed at $0.01$.
As shown in Fig.~\ref{fig:4}~(b), the effective $g$-factor decreases with the increase of
$T_{\rm K}$.
It becomes smaller for each $T_{\rm K}$ as $\Delta_{\downarrow \downarrow}$ is
decreased from $\Delta_{\uparrow \uparrow}$.
As long as we see Fig.~\ref{fig:4}(b), the data for
$\Delta_{\downarrow \downarrow} = 0.001$ ($10 \%$ of $\Delta_{\uparrow \uparrow}$)
is not much different from those for $\Delta_{\downarrow \downarrow} = 0$ where
infinitesimal excitations are dominant.
Therefore the spin polarization of the energy gap just modifies the effective
$g$-factor even if one of the spin dependent order parameters vanishes.
The local spin is completely quenched only when both $\Delta_{\uparrow \uparrow}$
and $\Delta_{\downarrow \downarrow}$ vanish.
\par

The $\Delta_{\downarrow \downarrow} = 0$ state mentioned above has infinite low-energy
excitations below the energy gap associated with only down spin of the conduction electrons.
This completely polarized state is described by the following NRG Hamiltonian:
\begin{eqnarray}
&& H_N^* = H_N (\tilde{J} = 0) - J^* \Lambda^{(N-1)/2}
f_{0 \downarrow}^{\dagger}f_{0 \downarrow} S_z
\nonumber \\
&&~~~~ + {h^* \over 2} [1 + (-1)^N] S_z,
\end{eqnarray}
where $J^*$ is an effective coupling constant and $h^*$ is an effective magnetic
field, which are determined by the low-lying excitation energies.
The latter is an effect of the Cooper pairs with only up spin.
The NRG Hamiltonian $H_N (\tilde{J} = 0)$ for the free electrons
can be described with quasiparticle and quasihole excitations as
\begin{eqnarray}
&& H_N (\tilde{J} = 0) = \sum_{i = 1}^{(N+1)/2} \eta_i
(g_i^{\dagger} g_i + h_i^{\dagger} h_i) ~~ (N{\rm : odd}), \\
&& H_N (\tilde{J} = 0) = \eta'_0 g_0^{\dagger} g_0
\nonumber \\
&&~~~~~~~~~~~~~~~ + \sum_{i = 1}^{N/2} \eta'_i
(g_i^{\dagger} g_i + h_i^{\dagger} h_i) ~~ (N{\rm : even}),
\end{eqnarray}
where $g_i$ and $h_i$ are annihilation operators for the particles and holes, respectively.
The energies for $\Lambda = 3$ are given by $\eta_1 = 0.800$, $\eta_2 = 2.997$, $\cdots$,
$\eta_i = 3^{i-1}$, and $\eta'_0 = 0$, $\eta'_1 = 1.696$, $\eta'_2 = 5.196$, $\cdots$,
$\eta'_i = 3^{i-1/2}$.
The fermion operator at the impurity site is given by
\begin{eqnarray}
&& f_{0 \downarrow} = 3^{-(N-1)/4} \sum_{i=1}^{(N+1)/2}
\alpha_i (g_i + h_i^{\dagger}) ~~ (N{\rm : odd}), \\
&& f_{0 \downarrow} = 3^{-(N-1)/4} [\alpha'_0 g_0 
\nonumber \\
&&~~~~~+ \sum_{i=1}^{N/2}
\alpha'_i (g_i + h_i^{\dagger})] ~~ (N{\rm : even}),
\end{eqnarray}
where the coefficients for $\Lambda = 3$ are $\alpha_1 = 0.628$, $\alpha_2 = 0.895$, $\cdots$,
$\alpha_i = \alpha 3^{(i-1)/2}$ ($\alpha = 0.5155$), and
$\alpha'_0 = 0.620$, $\alpha'_1 = 0.706$, $\cdots$, $\alpha'_i = \alpha' 3^{(i-1)/2}$
($\alpha' = 0.6784)$.
Then the effective NRG Hamiltonian for describing the case of $\Delta_{\downarrow \downarrow} = 0$
is finally expressed with
\begin{eqnarray}
&& H_N^* = \sum_{i=1}^{(N+1)/2} \eta_i (g_i^{\dagger} g_i + h_i^{\dagger} h_i)
\nonumber \\
&&~~~~~ - J^* \sum_i \alpha_i^2 (g_i^{\dagger} g_i - h_i^{\dagger} h_i) S_z
\mbox{} + O([J^*]^2),
\end{eqnarray}
for odd numbers of the renormalization step $N$ and
\begin{eqnarray}
&& H_N^* = \eta'_0 g_0^{\dagger} g_0
\mbox{} + \sum_{i=1}^{N/2} \eta'_i (g_i^{\dagger} g_i + h_i^{\dagger} h_i)
\nonumber \\
&&~~~~~ -J^* (\alpha'_0)^2 g_0^{\dagger} g_0 S_z 
\mbox{} - J^* \sum_i (\alpha'_i)^2 (g_i^{\dagger} g_i - h_i^{\dagger} h_i) S_z
\nonumber \\
&&~~~~~ + h^* S_z + O([J^*]^2),
\label{eqn:18}
\end{eqnarray}
for even numbers.
In order to reproduce the low-energy excitations, $J^*$ and $h^*$ in eq.~(\ref{eqn:18})
must satisfy the following relation:
\begin{equation}
\eta'_0 - {J^* \over 2} (\alpha'_0)^2 + {h^* \over 2} = - {h^* \over 2},
\end{equation}
which gives a twofold degenerate ground state ($| - \rangle$ and $g_0^{\dagger} | + \rangle$),
where $|\pm \rangle$ represents the $S_z = \pm 1/2$ local spin state.
Due to the effective field $h^* = (J^* / 2) (\alpha'_0)^2$, all the NRG low-energy excitations
are doubly degenerate as obtained for odd numbers of $N$.
\par

If the triplet superconducting state is realized, the spin-polarized excitations
are dominant below the higher transition temperature associated with a larger one
of $|\Delta_{\uparrow \uparrow}|$ and $|\Delta_{\downarrow \downarrow}|$.
In the temperature region between the two transition temperatures, the local spin
behaves like a classical spin and it is coupled to the spin-polarized conduction electrons.
When the second transition occurs at the lower temperatures, the energy gap opens
completely and the Kondo effect generates a bound state.
\par

\section{Conclusion}
This study shows a variety of Kondo effect influenced by the triplet Cooper pairs.
The Kondo effect is affected not only by the energy gap but also
by the orbital degrees of freedom of the Cooper pairs, which is completely different
from the $s$-wave case.
Due to the spin doublet ground state, $\chi_{\rm local}$ displays the Curie law at sufficiently
low temperatures, and the effective $g$-factor of the local spin is strongly reduced
in the large $T_{\rm K} /\Delta$ region.
The Curie term $T \chi_{\rm imp} = 1/4$ indicates that the spin of the ground state is generated
by the $l=1$ conduction electrons via the orbital effect of the $p_x +{\rm i} p_y$-wave
Cooper pairs.
In addition, the spin-polarization effect is unique to such triplet
superconducting states.
If either $\Delta_{\uparrow \uparrow}$ or $\Delta_{\downarrow \downarrow}$ vanishes,
the local spin is not quenched by the Kondo effect and behaves like a classical spin.
\par

All of the results are useful for explaining
the coexistence of magnetism and such unconventional superconductivity.
This argument is valid for heavy fermion superconductors realizing
$T_{\rm K} / \Delta \gg 1$, while in the same condition the local moment vanishes
at low temperatures for the $s$-wave superconductivity.
\par

\section*{Acknowledgements}
We would like to thank O. Sakai and Y. Shiina for useful discussions.
This work is supported by JSPS for Encouragement of Young Scientists (No.~13740214).



\end{document}